\documentclass{PoS}

\def\beq{\begin{equation}}
\def\eeq{\end{equation}}
\def\bea{\begin{eqnarray}}
\def\eea{\end{eqnarray}}
\def\beqn{\begin{eqnarray}} \def\eeqn{\end{eqnarray}}
\def\beeq{\begin{eqnarray}}
\def\eeeq{\end{eqnarray}}

\def\ep{\epsilon}

\def\nn{\nonumber}
\def\Eq#1{Eq.~(\ref{#1})}

\def\td#1{\tilde{\delta}\left(#1\right)}

\newcommand{\la}{\langle}
\newcommand{\ra}{\rangle}

\def\M#1{{\cal M}^{(#1)}}

\def\qq{{q \bar q}}
\def\qqg{{q \bar q g}}

\newcommand\as{\alpha_{\mathrm{S}}}

\def\as{\alpha_{\rm S}}

\def\v{{\rm V}}
\def\r{{\rm R}}

\title{
\vspace*{-1.9cm}
\begin{minipage}{\textwidth}
{\normalfont\small IFIC/16-81
\hspace{\fill} November 2016
}\\
\end{minipage}\\[60pt]
Towards regularized higher-order computations in QFT without DREG}

\ShortTitle{Towards regularized higher-order computations in QFT without DREG}

\author{\speaker{G. F. R. Sborlini}$^{\ a,b}$, F. Driencourt-Mangin$^a$, R. J. Hern\'andez-Pinto$^c$ and G. Rodrigo$^a$\\\\
        $^a$Instituto de F\'{\i}sica Corpuscular, Universitat de Val\`{e}ncia -- 
Consejo Superior de Investigaciones Cient\'{\i}ficas, Parc Cient\'{\i}fic, E-46980 Paterna, Valencia, Spain.\\\
        $^b$Dipartimento di Fisica, Universit\`a di Milano and INFN Sezione di Milano,
I-20133 Milan, Italy.\\\
        $^c$Facultad de Ciencias F\'isico-Matem\'aticas, Universidad Aut\'onoma de Sinaloa, Ciudad Universitaria, CP 80000, Culiac\'an, Sinaloa, M\'exico.\\\\
        E-mail: \email{german.sborlini@ific.uv.es, felix.dm@ific.uv.es,roger@uas.edu.mx,german.rodrigo@csic.es}}

\abstract{In this talk, we review the basis of the loop-tree duality theorem, which allows to rewrite loop scattering amplitudes in terms of tree-level like objects. Since the loop measure is converted into a phase-space one, both virtual and real contributions are expressible using the same integration variables. A physically motivated momentum mapping allows to generate the real emission process starting from the Born kinematics and the loop momenta. The integrand-level combination leads to regular functions, which can be integrated without using dimensional regularization (DREG) and correctly reproduce the finite higher-order corrections to physical observables. We explain the implementation of this novel approach to compute some benchmark physical processes, and we show how to deal with both infrared and ultraviolet divergences in four space-time dimensions.}

\FullConference{38th International Conference on High Energy Physics\\
                  3-10 August 2016\\
                 Chicago, USA}

\begin{document}

\section{Introduction}
\label{sec:sec1}
In the context of the very precise experimental measurements coming from the LHC, the theoretical predictions are forced to fulfil increasing accuracy requirements. For this reason, the high-energy physics community is unveiling a great effort to compute higher-orders in perturbative QCD, as well as EW corrections which were considered subleading long time ago.

In order to achieve this purpose it is mandatory to properly (and efficiently) deal with divergent expressions in intermediate steps of the computation. This requires using a regularization method such as dimensional regularization (DREG) \cite{Bollini:1972ui,'tHooft:1972fi,Cicuta:1972jf,Ashmore:1972uj}, which extends the dimension of the space-time from $d=4$ to $d=4-2\ep$. In this way, DREG transforms divergences or singularities into $\epsilon$-poles. After the cancellation of ultraviolet (UV) singularities through the application of a well-established renormalization program, the infrared (IR) singularities must still be removed by a proper combination of the real-radiation with the virtual contributions. According to the Kinoshita-Lee-Nauenberg (KLN) theorem \cite{Kinoshita:1962ur,Lee:1964is}, if the observable under consideration is IR-safe, then the real-virtual combination lacks of IR singularities. Moreover, within DREG, this implies that the $\epsilon$-poles in the renormalized virtual part cancel exactly those present in the real-radiation contribution. However, it is worth appreciating that even if the final result is finite, it is not possible to straightforwardly remove the regulator in intermediate steps. Explicitly, within DREG, the limit $\ep \to 0$ does not commute with the integration symbol. 

The last observation might seem to be a technical detail, but keeping track of the regulator across the whole computation could lead to very complicated expressions. This translates directly into a noticeable increase in the computational complexity, which makes a really hard-task to tackle multi-leg processes at NNLO or beyond. For instance, in the framework of the subtraction method \cite{Kunszt:1992tn,Frixione:1995ms,Catani:1996jh,Catani:1996vz}, analytical integration is required to isolate the $\epsilon$-poles and, then, local counter-terms must be added to the real contributions before any attempt of numerical calculation. To expose the key points of this well-established formalism, let's consider a NLO cross-section for a $2\to m$ scattering process. Since the total cross-section is the most inclusive observable, it is IR-safe and KLN theorem applies. Thus, the IR singularities present in the virtual (i.e. one-loop amplitudes) must be cancelled with the phase-space (PS) integral of the real-radiation term, which corresponds to a $2 \to m+1$ process. Rephrasing this sentence, the loop-integration and the PS integral over the extra-real particle contain \emph{exactly} the same IR divergences (with opposite sign). For this reason, a counter-term can be build in such a way that it locally cancels the divergences in the $m+1$ PS and its integrated form contains the same $\ep$-poles originated in the virtual. Explicitly,
\beqn
d \sigma^{\rm NLO} &=& \left[\,\int_m \, \left(d\sigma_{\rm V}-\int_1 \, dR \right)\,\right]_{\ep=0} \,  + \int_{m+1} \left[d\sigma_{\rm R}+dR \right]_{\ep=0}\, \, ,
\label{eq:SUBTRACTIONMAIN}
\eeqn
with $dR$ the differential form of the counter-terms. The different variations of the subtraction method provide alternative paths to build these terms, although the core idea remains the same: a partial cancellation of singularities \emph{after} integration and local counter-terms to render the real-radiation term integrable. In any case, it is worth appreciating that, in general, the limit $\ep \to 0$ can be taken before integration only in the second term of \Eq{eq:SUBTRACTIONMAIN}.

The idea of our work is to exploit the underlying origin of the IR singularities, and use directly the real-emission amplitude as a counter-term for the renormalized virtual-contribution. In this way, we avoid introducing IR counter-terms that have to be added/subtracted and integrated using regulators. Moreover, as we will explain later, the local cancellation of singularities is done in such a way that the $\ep \to 0$ limit can be safely considered at integrand level, i.e. \emph{before} integration. In consequence, the method results in a purely algebraic algorithm that does not need any additional regularization, since the emerging integrands are naturally integrable functions. This motivates the denomination of four-dimensional unsubtraction (FDU) approach \cite{Hernandez-Pinto:2015ysa,Sborlini:2015uia,Sborlini:2016fcj,Sborlini:2016gbr,Rodrigo:2016hqc,Sborlini:2016hat,Hernandez-Pinto:2016uwx}, as we will explain in the forthcoming sections.

\section{Four-dimensional unsubtraction (FDU)}
\label{sec:sec2}
The central idea of the FDU approach is to combine real and virtual contributions before integration, expressing virtual amplitudes as phase-space integrals, and relocating the IR singularities in the real-emission domain through a suitable change of variables. On top of that, it is necessary to deal with unintegrated renormalization counter-terms\footnote{We refer the interested reader to Refs. \cite{Hernandez-Pinto:2015ysa,Sborlini:2016gbr,Sborlini:2016hat}, where an extensive discussion about this topic is presented.}. In the following discussion, we will explain explicitly how to deal with the method at NLO, although it can be generalized to the multi-loop case.

There are two key components behind the FDU method:
\begin{enumerate}
\item the loop-tree duality theorem (LTD) \cite{Catani:2008xa,Bierenbaum:2010cy},
\item a universal physically-motivated momentum mapping.
\end{enumerate}
Briefly speaking, the LTD theorem establishes that virtual amplitudes can be expressed as a sum of single cuts. It is closely related with the Feynman tree-theorem, which reduces loops to a sum over all possible multiple-cuts. Both approaches are equivalent \cite{Catani:2008xa} because LTD introduces a modified prescription which takes into account the correlations induced beyond the single cuts. To be more concrete, let's consider an $m$-particle scalar one-loop Feynman integral. If external particles are regarded as outgoing, with momentum $p_i^{\mu}$ and the loop-momentum flow is counter-clockwise, then we define the internal momenta as $q_i^{\mu}=\ell^{\mu} + p_1^{\mu} + \ldots + p_i^{\mu}$, with $q_m^{\mu}=\ell^{\mu}$ because of momentum conservation. Then, LTD implies that
\beqn
\int_{\ell}\, \prod_{i=1}^m\,\frac{1}{q_i^2-m_i^2+\imath 0} &=& -\sum_{i=1}^{m} \, \int_{\ell} \td{q_i} \, \prod_{j=1,i\neq j}^{m} G_D(q_i;q_j) \, ,
\label{eq:DualExpressionONELOOP}
\eeqn
where $G_D(q_i;q_j)$ is the dual-propagator, with the prescription transformation $+\imath 0 \to - \imath 0 \, \eta\cdot(q_j-q_i)$. It is worth appreciating that this formulae is valid under the assumption of single powers of the propagators. Otherwise, we must compute the residue by making use of Cauchy's formulae for higher-order poles, as carefully explained in Refs. \cite{Bierenbaum:2012th}. On the other hand, notice that the r.h.s. of \Eq{eq:DualExpressionONELOOP} contains a sum of \emph{PS-like integrals} since $q_i$ is forced to be on-shell and acts as if it were an additional particle. This fact is crucial to achieve a regular (i.e. free of singularities) integrand-level real-virtual combination. 

\begin{figure}[ht]
\begin{center}
\includegraphics[width=0.74\textwidth]{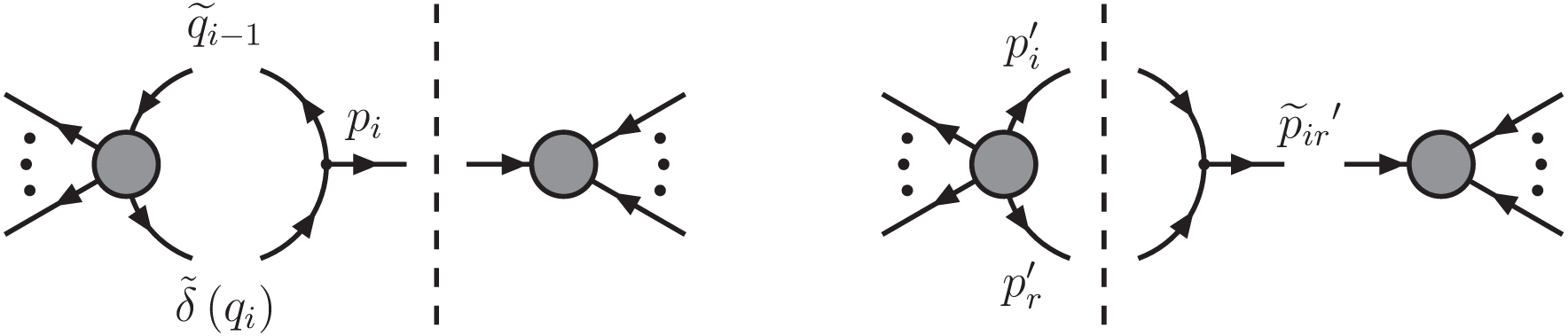}
\caption{\label{fig:FACTORIZACION}
Diagramatic contributions in the collinear limit, for both the dual one-loop (left) and real-emission tree-level squared amplitudes (right). The lines that are crossed by a dashed line correspond to on-shell states. When particles are collinear, the parent becomes on-shell and the diagram factorises.}
\end{center}
\end{figure}

Once LTD is applied to the virtual amplitude, we obtain a set of dual contributions. From them we can extract useful information about the location of the singularities, and the components that originate them. As explained in Refs. \cite{Buchta:2014dfa,Buchta:2015jea,Buchta:2015wna}, the intersection of forward and backward hyperboloids defined by the on-shell condition allows to identify the IR (and threshold) singularities. Moreover, this approach leads to proof the compactness of the region developing IR divergences \cite{Hernandez-Pinto:2015ysa,Sborlini:2016gbr,Sborlini:2016hat}. In fact this is crucial because the real-emission PS is finite and it contains all the IR-singularities that manifest in the virtual amplitudes. So, solving the real-virtual cancellation of singularities is equivalent to properly mapping the singular points in two compacts spaces. 

The underlying concepts to connect Born kinematics ($m$-particle PS) with the real-emission kinematics ($m+1$-particle PS) are rather similar to those used in the context of the dipole method~\cite{Catani:1996jh,Catani:1996vz}. After obtaining the dual amplitudes, we are left with $m$ external momenta and a free on-shell loop momentum; thus, dual amplitudes contain an \emph{extra} on-shell momentum. This is exactly what we have in the real contribution: $m+1$ on-shell momenta in the final state. The next step consists in isolating the singularities in the $m+1$ PS. Explicitly, if $p_i'^{\mu}$ are the momenta of the real-emission partons, let's define the partition
\beq
{\cal R}_i = \{{y'}_{i r} < {\rm min}\, {y'}_{jk} \} \, , \quad \quad \sum_{i=1}^{m} {\cal R}_i =1 \, , 
\eeq
where ${y'}_{ij}=2\, p_i' \cdot p_j'/Q^2$, $r$ is the radiated parton from parton $i$, and $Q$ is the typical hard scale of the scattering process. Thus, the only allowed collinear/soft configurations in ${\cal R}_i$ are $i\parallel r$ or $p_r'^{\mu} \to 0$. On the other hand, there are $m$ dual contributions, each one associated with a single cut of an internal line. So, we can establish an identification among partitions and dual amplitudes, based in the picture shown in Fig.~\ref{fig:FACTORIZACION}. In concrete, the cut-line in the dual amplitude must be interpreted as the extra-radiated particle in the real contribution; i.e. $q_i \leftrightarrow p_r'$. Then, we settle in one of the partitions, for instance ${\cal R}_i$. Because the only collinear singularity allowed is originated by $i \parallel r$, we distinguish particle $i$ and call it the {\it emitter}. After that, we single out all the squared-amplitude-level diagrams in the real contribution that become singular when $i \parallel r$ and cut the line $i$. These have to be topologically compared with the dual-Born interference diagrams whose internal momenta $q_i$ are on-shell (i.e. the line $i$ is cut), as suggested in Fig. \ref{fig:FACTORIZACION}. Summarizing this procedure: the dual contribution $i$ is to be combined with the real-contribution coming from region ${\cal R}_i$.

The momentum mapping required is motivated by the general factorisation properties in QCD \cite{Buchta:2014dfa,Catani:2011st} and the topological identification in Fig.~\ref{fig:FACTORIZACION}. Explicitly, let's take the $m+1$-particle real-emission kinematics, with $i$ as the emitter and $r$ as the radiated particle, and we introduce a reference momentum, associated to the spectator $j$. The multi-leg momentum mapping is given by 
\bea
&& p_r'^\mu = q_i^\mu~, \ \ \quad \quad \quad \quad \quad \quad \quad p_j'^\mu = (1-\alpha_i) \, p_j^\mu~, \nn \\
&& p_i'^\mu = p_i^\mu - q_i^\mu + \alpha_i \, p_j^\mu~, \qquad \alpha_i = \frac{(q_i-p_i)^2}{2 p_j\cdot(q_i-p_i)}~,
\label{multilegmomentummapping}
\eea
where $p_l'^\mu$ denotes the momenta in the real-emission process ($p_k'^2=0$ and $\sum_l \, {p'}_l =0$, in the massless case). This transformation does not alter the initial-state momenta ($p_a$ and $p_b$) neither $p'_k$ with $k \ne i,j$, and momentum conservation is respected since $p_i+ p_j +\sum_{k\ne i,j} p_k = p_i'+p_r'+p_j'+\sum_{k\ne i,j} p_k'$. 

\section{Application example: $\gamma^* \to q \bar q$ at NLO}
\label{sec:sec3}
\begin{figure}[ht]
\begin{center}
\includegraphics[width=0.9\textwidth]{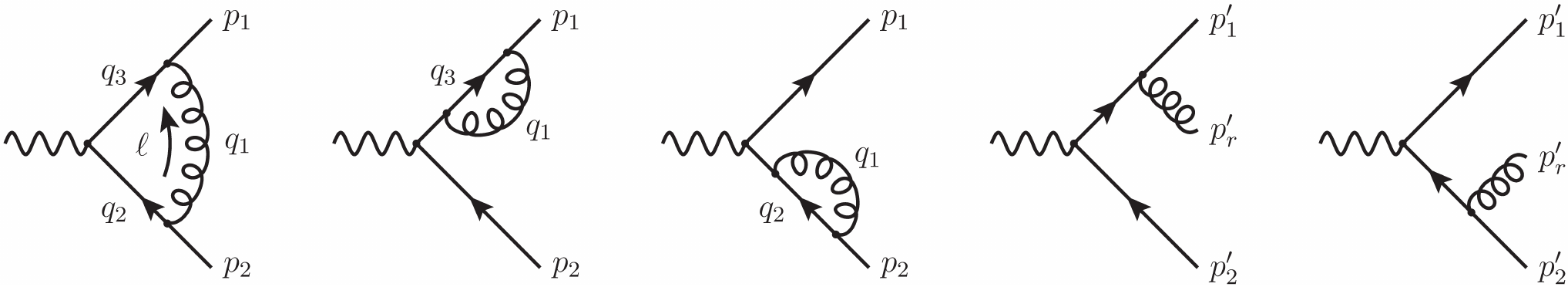}
\caption{\label{Diagramas}
Feynman diagrams contributing to the NLO QCD corrections to $\gamma^* \to q \bar q (+ g)$, with massless quarks. Even if they vanish after integration in DREG, self-energy diagrams contribute in a non-trivial way at integrand level, and they are a crucial point to cancel IR singularities present in the real-radiation PS integrals.}
\end{center}
\end{figure}

In the first place, we consider the complete set of ${\cal O}(\as^2)$ diagrams, for both the real and the virtual contributions as depicted in Fig. \ref{Diagramas}. We define the total \emph{unrenormalized} virtual cross-section as
\beq
\sigma_{\v}^{(1)} = \frac{1}{2s_{12}} \, \int d\Phi_{1\to 2} \, \left( 2 {\rm Re}
\la \M{0}_\qq| \M{1}_\qq \ra + (\Delta Z_2(p_1)+\Delta Z_2(p_2))\, |\M{0}_\qq|^2\right)~,
\label{eq:virtualunrenormalized}
\eeq
where we distinguished the contributions originated in the triangle diagram from those related to the self-energies\footnote{The self-energy diagrams have to be properly rewritten to imitate the IR singular behaviour of the corresponding real-emission counter-parts. For more details about this relevant observation, see Refs. \cite{Sborlini:2016gbr,Sborlini:2016hat,2016PROCMASSIVE}.}. After that, we must introduce the local UV counter-terms which implements the desired renormalization scheme; for instance, in Ref. \cite{Sborlini:2016gbr} we obtained the explicit formulae for the $\overline{\rm MS}$ scheme. Then, we apply LTD to \Eq{eq:virtualunrenormalized} in order to obtain the corresponding three dual contributions, $\widetilde \sigma_{\v,i}^{(1)}$.

As explained in Sec. \ref{sec:sec2}, a partition of the real-emission phase-space must be introduced, thus leading to
\beq
\widetilde \sigma_{\r,i}^{(1)} = \frac{1}{2s_{12}} \, \int d\Phi_{1\to 3} \,  |{\cal M}_{\qqg}^{(0)}|^2 \, \theta(y_{jr}' -y_{ir}' )~
\qquad i,j \in \{ 1,2\} \, , \quad i \neq j \ \ ,
\label{realalqqg}
\eeq
which fulfils $\widetilde \sigma_{\r,1}^{(1)}+\widetilde \sigma_{\r,2}^{(1)} = \sigma_{\r}^{(1)}$. After that, we apply the real-virtual mapping in each partition. In this way, the cross-cancellation of singularities among the dual contributions will take place directly in four dimensions. 

Finally, we get the master formula
\beq
\sigma^{(1)} = {\cal T}\left(\sum_{i=1}^3 \, \widetilde \sigma_{\v,i}^{(1)}+\, \sum_{j=1}^2 \, \widetilde \sigma_{\r,j}^{(1)}\right) - \widetilde \sigma^{(1)}_{\rm UV} \, ,
\label{MasterFORMULA}
\eeq
where $\widetilde \sigma^{(1)}_{\rm UV}$ is the dual representation of the local UV counter-terms and ${\cal T}$ is an operator that implements the unification of dual-coordinates at integrand level (with the corresponding Jacobians). If we add all the contributions at integrand level and deal with a single master integration, the expression in \Eq{MasterFORMULA} is directly implementable in four dimensions and leads to the right result after numerical computation. It is worth mentioning that, in order to improve the numerical stability, it is requested to compactify the integration domain, applying a transformation as suggested in Ref.~\cite{Sborlini:2016hat}. 

\section{Conclusions and outlook}
\label{sec:sec4}
The four-dimensional unsubtraction approach is a novel technique, based on the LTD theorem, that algorithmically leads to a numerical implementation of higher-order computations. The underlying idea is that both the real and the virtual contributions, for any IR-safe observable, share the same divergent behaviour, even though they are expressed by using different variables. With the application of LTD, the loop amplitude is rewritten as a PS-like integral, and the integration domain is expressible with the same variables used for the real-emission contribution. After that, a proper momentum-mapping, physically motivated by a diagram-by-diagram analysis, allows to generate the real-emission kinematics starting from the Born process and the loop three-momentum. Moreover, this transformation maps the IR singularities to the same points in the integration domain, thus rendering the real-virtual combination integrable in four dimensions. Additionally, the UV behaviour must be locally regularized by adding local renormalization counter-terms. It is important to emphasize that these counter-terms, as well as self-energy contributions, have to be taken into account even if they vanish within DREG.

In this article we focus in the general aspects of the method, and center in the massless case. The algorithm can be generalized to deal with massive particles, as we describe in Ref. \cite{Sborlini:2016hat,2016PROCMASSIVE}. Beyond that, this approach could be extended to deal with multi-loop multi-leg processes, therefore offering an interesting alternative to the standard methods available in the HEP community.

\section*{Acknowledgments}
This work is partially supported by the Spanish Government, EU ERDF funds (grants FPA2014-53631-C2-1-P and SEV-2014-0398) and by GV (PROMETEU II/2013/007).

\end{document}